\documentclass[12pt,a4paper,twocolumn,fleqn]{narms}

\usepackage{subfigure}
\usepackage{epsfig,color}
\usepackage{timesmt}
\usepackage{amsmath}
\usepackage{amssymb}
\usepackage{url}
\usepackage{hyperref}

\usepackage{chicaco}

\begin{document}
\title{Efficient variance-based reliability sensitivity analysis\\ for Monte Carlo methods}
\author{{T. Most} \\
{\aff{Institute of Structural Mechanics, Bauhaus-University, Weimar, Germany}}}

\date{}

\maketitle

\section{INTRODUCTION}
In the application of state-of-the-art reliability methods, often the quantification of the influence of the stochastic parameter properties with respect to the estimated failure probability is necessary. In an iterative improvement of the investigated structure or system, this is usually done within the reliability-based design approach. In the framework of an uncertainty quantification, often classical variance-based sensitivity measures are used to quantify the global influence of the input variation to the variation of the interesting output quantities. This global variance-based approach gives a suitable estimate of the input variable influence on the computed failure probability, as long as the investigated limit state function is almost linear and the input variables are almost normally distributed. 

For strongly nonlinear examples e.g. the investigation of stability problems of engineering structures, the classical global variance-based approach might lead to a wrong quantification of the input variable importance with respect to the failure probability, especially in case of multiple failure regions.
In this paper, a Monte Carlo based approach for the quantification of the importance of the scattering input parameters with respect to the failure probability is presented. Using the basic idea of the $\alpha$-factors of the First Order Reliability Method \shortcite{Papaioannou2021}, this approach was developed to analyze arbitrary marginal parameter distributions. 
Based on an efficient transformation scheme using the importance sampling principle, only a single analysis run by a plain or variance-reduced Monte Carlo method is required to give a sufficient estimate of the introduced parameter sensitivities. 

As basic idea, the derivatives of the failure probability with respect to the input parameter variance, are used to quantify the importance of the specific input parameters. This assumptions results in similar results as for the classical variance-based measures if the linear limit state function is linear and the input parameters are normal. In cases of non-linear limit states or non-normal input parameters, the suggested sensitivity measures are much more suitable for deciding upon the parameter importance with respect to the investigated reliability problem as the classical variance-based measures.

\section{Definition of uncertainties}
In our study, we assume scalar random parameters which are assembled in a random vector 
\begin{equation}
\mathbf{X}=\left[ X_1, X_2, \ldots, X_m\right],
\end{equation}
which may contain continuous and discrete random numbers $X_i$.

The marginal distributions of the individual random numbers can be represented by standard distribution types, such as normal, log-normal, truncated normal and uniform types as well as more flexible types, such as the beta distribution, the generalized lamba distribution \shortcite{Karian2000} and a piecewise-uniform distribution function.

For normally-distributed variables a linear dependence between the random input parameters can be represented with the Gaussian copula in closed form
\begin{equation}
f_\mathbf{X}(\mathbf{x}) = \frac{1}{\sqrt{(2\pi)^m |\mathbf{C}_\mathbf{XX}|}}
\exp\left[-\frac{1}{2}(\mathbf{x} - \mathbf{\bar X})^T\mathbf{C}_\mathbf{XX}^{-1}(\mathbf{x} - \mathbf{\bar X})\right],
\label{normal_density}
\end{equation} 
where $f_\mathbf{X}(\mathbf{x})$ is the joint probability density function of random vector $\mathbf{X}$ and $\mathbf{\bar X}$ is the corresponding mean vector and $\mathbf{C}_\mathbf{XX}$ the covariance matrix.
In our study, the Nataf model \shortcite{Nataf1962} is used to extend the Gaussian correlation model to non-Gaussian distribution types. 
In the Nataf approach the marginal distributions of the random variables are transformed to the standard Gaussian distribution. In this transformed space, a Gaussian copula as given in equation~\ref{normal_density} is assumed. The correlation coefficients of the standardized Gaussian space are obtained from the correlation coefficients of the original distributions by an iterative procedure \shortcite{Bucher2009_Book} or analytical regression function \shortcite{Liu1986}. 

\section{Variance based sensitivity analysis}

\subsection{First order and total effect sensitivity indices}
Assuming a model with a scalar output $Y$ as a function of a given set of $m$ random input variables $X_i$
\begin{equation}
Y=f(X_1,X_2,\ldots,X_m),
\label{model}
\end{equation}
the first order sensitivity measure was introduced as \shortcite{Sobol1993}
\begin{equation}
S_{X_i}=\frac{V_{X_i}(E_{\mathbf{X}_{\sim i}}(Y|X_i))}{V(Y)},
\label{first}
\end{equation}
where $V(Y)$ is the unconditional variance of the model output and 
$V_{X_i}(E_{\mathbf{X}_{\sim i}}(Y|X_i))$ is named the {\it variance of  conditional expectation}
with $\mathbf{X}_{\sim i}$ denoting the matrix of all factors but $X_i$.
$V_{X_i}(E_{\mathbf{X}_{\sim i}}(Y|X_i))$ measures the first order effect of $X_i$ on the model output.

Since first order sensitivity indices measure only the decoupled influence of each variable
an extension for higher order coupling terms was introduced.
The total effect sensitivity indices have been introduced \shortcite{Homma1996} as
\begin{equation}
S^T_{X_i}=1-\frac{V_{\mathbf{X}_{\sim i}}(E_{X_i}(Y|\mathbf{X}_{\sim i}))}{V(Y)},
\label{total}
\end{equation}
where $V_{\mathbf{X}_{\sim i}}(E_{X_i}(Y|\mathbf{X}_{\sim i}))$ measures the first order effect of $\mathbf{X}_{\sim i}$
on the model output which does not contain any effect corresponding to $X_i$.

The consideration of correlated input parameters in the context of the Nataf model has been presented in \shortcite{Most_2012_REC_Correlated}.

\subsection{Sensitivity indices of a linear model}
Assuming a linear model
\begin{equation}
Y = a_0 + \sum_{i=1}^m a_i \cdot X_i,
\end{equation}
the variance of the model response reads in case of independent random variables
\begin{equation}
\sigma_Y^2 = \sum_{i=1}^m a_i^2 \cdot \sigma_{X_i}^2.
\end{equation}
The first order sensitivity index for this linear model can be derived as
\begin{equation}
S^{lin}_{X_i} = \frac{a_i^2 \cdot \sigma_{X_i}^2}{\sum_{j=1}^m a_j^2 \cdot \sigma_{X_j}^2}, \quad\sum_{i=1}^m S^{lin}_{X_i} = 1.
\end{equation}
In this special case, higher order effects vanish and the first order index is equivalent to the total effect index for each input variable. The sum of the indices for all input variables is equal to one.

\section{Reliability-based sensitivity indices}
For a given set of jointly distributed random variables $X_i$ and a limit state function $g(\mathbf{X})$
the probability of failure $P_f$ can be determined via integration
\begin{equation}
\begin{aligned}
P_f &= P\left[\mathbf{X} : g(\mathbf{X})\leq 0\right]\\[2mm]
&= \idotsint\limits_{g(\mathbf{X})\leq 0} f_\mathbf{X}(\mathbf{x})d\mathbf{x}.
\end{aligned}
\label{failure_probability}
\end{equation}
The limit state function divides this random variable space into a safe domain $S=\{\mathbf{x}:g(\mathbf{x})> 0\}$ and a failure domain
$F=\{\mathbf{x}:g(\mathbf{x})\leq 0\}$.

The computational challenge in
determining the integral of Eq.~(\ref{failure_probability}) lies
in evaluating the limit state function $g (\mathbf{x})$ at a specific position $\mathbf{x}$,
which for non-linear systems usually requires an
incremental/iterative numerical approach. 

\subsection{First Order Reliability Method}
Assuming a linear limit state function $g(\mathbf{x})$ of the basic variable vector $\mathbf{X}$
\begin{equation}
g(\mathbf{x}) = a_0 + \sum_{i=1}^m a_i \cdot x_i,
\end{equation}
with $m$ independent normally distributed input variables
\begin{equation}
X_i \sim N(\bar X_i, \sigma_{X_i}^2),
\end{equation}
the mean value and the variance of the limit state function can be derived as
\begin{equation}
\bar g = a_0 + \sum_{i=1}^m a_i \cdot \bar X_i, \quad \sigma^2_g = \sum_{i=1}^m a_i^2 \cdot \sigma_{X_i}^2.
\end{equation}
Using the First Order Reliability Method (FORM) the reliability index $\beta$ and the failure probability $P_f$
can be obtained as
\begin{equation}
\beta = \frac{\bar g}{\sigma_g} = \frac{a_0 + \sum a_i \cdot \bar X_i}{\sqrt{\sum a_i^2 \cdot \sigma^2_{X_i}}}, \quad P_f = \Phi(-\beta),
\label{form_linear}
\end{equation}
where the $\Phi$ denotes the cumulative distribution of a standard normal variable.

The derivative of the reliability index and the failure probability with respect to the variance $\sigma_{X_i}^2$ of each input variable can be derived according to \shortcite{Lu2008} as
\begin{equation}
\begin{aligned}
\frac{\partial \beta}{\partial (\sigma_{X_i}^2)} &= -\frac{1}{2}\bar g \cdot (\sigma_g^2)^{-\frac{3}{2}}  \cdot 
\frac{\partial (\sigma_g^2)}{\partial (\sigma_{X_i}^2)}\\
&= - \beta \frac{a_i^2}{\sum a_j^2 \cdot \sigma^2_{X_j}},\\
\frac{\partial P_f}{\partial (\sigma_{X_i}^2)} &= \frac{\partial P_f}{\partial \beta}\frac{\partial \beta}{\partial (\sigma_{X_i}^2)}\\
 &=  \frac{\beta}{2\pi} \cdot \exp\left(-\frac{\beta^2}{2}\right) \frac{a_i^2}{\sum a_j^2 \cdot \sigma^2_{X_j}}.
\end{aligned}
\end{equation}

If we now standardize the random variables to zero mean and unit standard deviation
\begin{equation}
X_i = U_i \cdot \sigma_{X_i} + \bar X_i,\quad U_i \sim N(0, 1) ,
\end{equation}
the limit state function reads 
\begin{equation}
g(\mathbf{u}) = \alpha_0 + \sum_{i=1}^m \alpha_i \cdot u_i,\\
\end{equation}
with 
\begin{equation}
\alpha_0 = a_0 + \sum_{i=1}^m a_i \cdot \bar X_i= \bar g, \quad \alpha_i = a_i \cdot \sigma_{X_i}.
\label{transform}
\end{equation}
In this equation $\alpha_i$ are the well known FORM-$\alpha$-values which define the position of the most probable failure point in the standard normal space.
The derivative of the failure probability with respect to the variance of the standard normal variables results 
\begin{equation}
\frac{\partial P_f}{\partial (\sigma_{U_i}^2)} =  \frac{\beta}{2\pi} \cdot \exp\left(-\frac{\beta^2}{2}\right)\frac{\alpha_i^2}{\sum \alpha_j^2}.
\end{equation}
If we standardize the derivatives with their sum, we obtain exactly the classical varianced-based sensitivity indices with respect to  standard normal variables
\begin{equation}
\frac{\partial P_f/\partial (\sigma_{U_i}^2)}{ \textstyle \sum_{j=1}^m \partial P_f/\partial (\sigma_{U_j}^2)} = \frac{\alpha_i^2}{\sum \alpha_j^2} = S^{lin}_{U_i}.
\end{equation}
Using equation~\ref{transform}, we can show that
\begin{equation}
S^{lin}_{X_i} = S^{lin}_{U_i} = \frac{\alpha_i^2}{\sum \alpha_j^2}.
\label{linear_indices}
\end{equation}

\subsection{Monte Carlo Simulation}
Based on the findings in the previous section, we define the varianced-based reliability sensitivity
as the normalized partial derivatives of the failure probability with respect to the variance of the input variables in the standard normal space
\begin{equation}
S^{rel}_{U_i} = \frac{\partial P_f/\partial (\sigma_{U_i}^2)}{ \textstyle \sum_{j=1}^m \partial P_f/\partial (\sigma_{U_j}^2)}.
\label{reli_index}
\end{equation}

In the Monte Carlo Simulation (MCS)\shortcite{Rubinstein1981}
the failure probability is estimated from a set of $N$ independent samples $\mathbf{x}_i$ as
\begin{equation}
\hat P_f=\frac{1}{N} \sum^N_{i=1} I\left(g(\mathbf{x}_i)\right),
\end{equation}
where the indicator function $I\left(g(\mathbf{x}_i)\right)$ is one if $g(\mathbf{x}_i)$ is negative or zero and zero else.

In our study, we define the reliability-based sensitivity indices according to equation~\ref{reli_index} as the standardized derivatives of the failure probability with respect to variance of the standard normal input variables. 
Assuming an important sampling principle, the samples of a modified joint probability density function can be scaled with the weightings to derived the failure probability with respect to the original probability density
\begin{equation}
\hat P_f=\frac{1}{N} \sum^N_{i=1} w(\mathbf{x}_i) I\left(g(\mathbf{x}_i)\right), \quad w(\mathbf{x}_i)=
\frac{f_{orig} (\mathbf{x}_i)}{f_{mod} (\mathbf{x}_i)},
\end{equation}
where $f_{orig} (\mathbf{x}_i)$ is the original joint probability density and $f_{mod} (\mathbf{x}_i)$ the modified density function with respect to the random vector $\mathbf{X}$, which was used to generate the samples $\mathbf{x}_i$.

For our study, we require the derivatives of the failure probability in the standard normal space according to equation~\ref{reli_index}.
These derivatives are calculated numerically by central differences but only the original Monte Carlo samples should be considered in this estimate and no additional samples should be generated. 
The initial Monte Carlo density is describted in the standard normal space as a jointly standard normal probability distribution with zero mean values
\begin{equation}
f_{\mathbf{U}_0}(\mathbf{u}) = \frac{1}{\sqrt{(2\pi)^m |\mathbf{C}_{\mathbf{UU}_0}|}}
\exp\left[-\frac{1}{2}\mathbf{u}^T\mathbf{C}_{\mathbf{UU}_0}^{-1}\mathbf{u}\right],
\label{normal_density2}
\end{equation} 
where $\mathbf{C}_{\mathbf{UU}_0}$ is the covariance matrix of the original distribution in the standard normal space
\begin{equation}
\mathbf{C}_{\mathbf{UU}_0} = \begin{pmatrix} 1 &\rho_{12}&\cdots &\rho_{1m}\\\\
\rho_{21}&1&\cdots &\rho_{2m}\\\\
\vdots &\vdots &\ddots &\vdots \\\\
\rho_{m1}&\rho_{m2}&\cdots &1\end{pmatrix},
\end{equation} 
which is the identity matrix in case of uncorrelated random inputs.

If we now apply a small change in the variance of a single input parameter $U_i$
\begin{equation}
(\sigma^2_{U_i})^\pm = \sigma^2_{U_i} \pm \Delta\sigma^2,
\end{equation} 
the corresponding derivates of the failure probability can be stimated
\begin{equation}
\begin{aligned}
\frac{\partial P_f}{\partial (\sigma_{U_i}^2)} &\approx \frac{1}{2N\Delta\sigma^2} \sum^N_{i=1} \left[w_{\Delta i}^+(\mathbf{u}_i) -w_{\Delta i}^-(\mathbf{u}_i)\right] I\left(g(\mathbf{x}_i)\right)\\
&=\frac{1}{2N\Delta\sigma^2} \sum^N_{i=1} \frac{f_{\Delta i}^+ (\mathbf{u}_i)-f_{\Delta i}^-(\mathbf{u}_i)}{f_{\mathbf{U}_0} (\mathbf{u}_i)} I\left(g(\mathbf{u}_i)\right),
\end{aligned}
\end{equation} 
where $f_{\Delta i}^+ (\mathbf{u}_i)$ is obtained by using $(\sigma^2_{U_i})^+$ and $f_{\Delta i}^- (\mathbf{u}_i)$ by $(\sigma^2_{U_i})^-$
as variance for the investigated input parameter $U_i$.
The modified probability density functions are calculated similar to equation~\ref{normal_density2} with zero mean but modified covariance matrix,
where the column and row of the investigated random variable $U_i$ are modified by a factor $d$
\begin{equation}
\mathbf{C}_{\mathbf{UU},{\Delta_i}} = 
\begin{pmatrix} 
1 &\rho_{12}&\cdots & d^\pm\rho_{1i} &\cdots &\rho_{1m}\\
\rho_{21}&1 &\cdots & d^\pm\rho_{2i} &\cdots  &\rho_{2m}\\
\vdots &\vdots & &\vdots & &\vdots \\
d^\pm\rho_{i1}&d^\pm\rho_{i2}&\cdots &(d^\pm)^2&\cdots &d^\pm\rho_{im}\\
\vdots &\vdots & &\vdots & &\vdots \\
\rho_{m1}&\rho_{m2}&\cdots&d^\pm\rho_{mi}&\cdots &1\end{pmatrix}
\label{covariance_delta}
\end{equation}
which is obtained from the defined central difference interval of the variance change
\begin{equation}
d^+ = \left( 1+\Delta\sigma^2\right)^{0.5},\quad d^- = \left( 1+\Delta\sigma^2\right)^{-0.5}.
\label{mcs_scaling_factor}
\end{equation}

Onces the derivative of the failure probability have been calculated for each input parameter $U_i$,
the sensitivity indices are obtained by simply normalizing with their sum according to equation~\ref{reli_index}.

\subsection{Importance Sampling}

The presented approach can be applied also for Importance Sampling,
where the importance sampling density is scaled for each parameter derivative with the same factors as given in 
equation~\ref{mcs_scaling_factor}.
The estimates of the partial derivatives read
\begin{equation}
\frac{\partial P_f}{\partial (\sigma_{U_i}^2)} \approx 
\frac{1}{2N\Delta\sigma^2} \sum^N_{i=1} \frac{f_{\Delta i}^+ (\mathbf{u}_i)-f_{\Delta i}^-(\mathbf{u}_i)}{f_{IS} (\mathbf{u}_i)}
 I\left(g(\mathbf{u}_i)\right), 
\end{equation} 
where $f_{\Delta i}^+ (\mathbf{u}_i)$ and $f_{\Delta i}^-(\mathbf{u}_i)$ are the  original density function modified with the central difference interval change in the covariance matrix according to equation~\ref{covariance_delta}, 
and $f_{IS} (\mathbf{u}_i)$ is the importance sampling density.
If we assume a single standard normal sampling density centered at the design point $\mathbf{u}_{IS}$ according to \shortcite{Bourgund1986},
the importance sampling density function reads
\begin{equation}
f_{IS} (\mathbf{u}_i) = \frac{\exp\left[-\frac{1}{2}(\mathbf{u}-\mathbf{u}_{IS})^T\mathbf{C}_{IS}^{-1}(\mathbf{u}-\mathbf{u}_{IS})\right]}{\sqrt{(2\pi)^m |\mathbf{C}_{IS}|}}.
\end{equation} 

In following analyses, we assume the importance sampling covariance as the identity matrix $\mathbf{C}_{IS}= \mathbf{I}$. However, the presented approach works as well for
an adapted sampling density as presented in \shortcite{Bucher1988}.

\section{Examples}
\subsection{Linear limit state function}
In the first example, we investigate a linear limit state function with five independent standard normal input parameters
\begin{equation}
g(\mathbf{x}) = b - 0.8 x_1 - 0.5 x_2 - 0.3 x_3 - 0.1 x_4 - 0.1 x_5,
\end{equation}
where the analytical solution results directly from equation~\ref{linear_indices} as
\begin{equation}
\begin{aligned}
S_{U_1} &= 0.64,&S_{U_2} &= 0.25,\\
S_{U_3} &= 0.09, &S_{U_4}&= S_{U_5} = 0.01.
\end{aligned}
\end{equation}
The reference solution for the reliability index is $\beta = b$ according to equation~\ref{form_linear}.

Different investigations have been carried out with Monte Carlo and Importance Sampling.
First the derivative step size for the central differences was investigated with the Monte Carlo Sampling and 10000 samples for $b=2.0$ which is equivalent to a failure probability of $0.023$. 100 indendent simulation runs have been carried out to estimate the mean value and the standard deviation of the sensitivity estimates.
\begin{table}[th]
\scriptsize
\begin{center}
\begin{tabular}{ccccccc}
\hline
& $\hat \beta$ & $\hat S_{U_1}$& $\hat S_{U_2}$& $\hat S_{U_3}$& $\hat S_{U_4}$& $\hat S_{U_5}$ \rule[0mm]{0pt}{2.5ex}\\[3pt]
   \hline
   Reference	& 2.000 &0.640 &0.250 &0.090 & 0.010 & 0.010  \rule[0mm]{0pt}{2.5ex} \\ 
   \hline
   $\Delta{\sigma^2} = 0.01 $	& &&&&   \rule[0mm]{0pt}{2.5ex} \\ 
	Mean values	&2.002& 0.635   & 0.247 &  0.083   & 0.020  &  0.014\\
	Std deviation	&0.026& 0.031  &  0.028  &  0.022 &   0.014  &  0.013\\
   \hline
   $\Delta{\sigma^2} = 0.05 $	& &&&&   \rule[0mm]{0pt}{2.5ex} \\ 
	Mean values	&2.002& 0.630  &  0.244  &  0.091  &  0.018 &   0.017\\
	Std deviation	&0.026& 0.033  &  0.028  &  0.024 &   0.014  &  0.014\\
   \hline
   $\Delta{\sigma^2} = 0.1 $	&& &&&   \rule[0mm]{0pt}{2.5ex} \\ 
	Mean values	&2.002& 0.632  &  0.246  &  0.088   & 0.016  &  0.017\\
	Std deviation	&0.026& 0.026  &  0.024  &  0.022 &   0.013  &  0.012\\
   \hline
   $\Delta{\sigma^2} = 0.2 $	&& &&&   \rule[0mm]{0pt}{2.5ex} \\ 
	Mean values	&2.002& 0.628  &  0.250  &  0.086  &  0.018  &  0.018\\
	Std deviation	&0.026& 0.029  &  0.026  &  0.026  &  0.014  &  0.015\\
   \hline
\end{tabular}
\end{center}
\caption{Linear example: influence of the numerical derivative step size on the accuracy of the MCS sensitivity indices}
\label{linear_mcs_results1}
\scriptsize
\begin{center}
\begin{tabular}{ccccccc}
\hline
& $\hat \beta$ & $\hat S_{U_1}$& $\hat S_{U_2}$& $\hat S_{U_3}$& $\hat S_{U_4}$& $\hat S_{U_5}$ \rule[0mm]{0pt}{2.5ex}\\[3pt]
   \hline
   Reference	& 2.000 &0.640 &0.250 &0.090 & 0.010 & 0.010  \rule[0mm]{0pt}{2.5ex} \\ 
   \hline
   $N = 1000$	& &&&&   \rule[0mm]{0pt}{2.5ex} \\ 
	Mean values	&2.003& 0.595 &   0.226 &    0.088 &    0.045 &   0.047\\
	Std deviation	&0.064& 0.087 &  0.078 & 0.062 & 0.033 & 0.036\\
   \hline
   $N = 10000$	& &&&&   \rule[0mm]{0pt}{2.5ex} \\ 
	Mean values	&2.002& 0.632 &   0.246  &   0.088 &  0.018  &    0.017\\
	Std deviation	&0.026& 0.026 &   0.024 &   0.022 &   0.013  &    0.012\\
   \hline
   $N = 100000$	& &&&&   \rule[0mm]{0pt}{2.5ex} \\ 
	Mean values	&2.000& 0.640 &  0.249 &   0.089 &  0.011 &   0.011\\
	Std deviation	&0.008& 0.009 &  0.008  &  0.007  & 0.006  &    0.006\\
   \hline
\end{tabular}
\end{center}
\caption{Linear example: influence of the sample size on the accuracy of the MCS sensitivity indices}
\label{linear_mcs_results2}
\end{table}
Table \ref{linear_mcs_results1} clearly indicates, that the estimates are very robust against the step size.
This is a result of the utilized central differences approach. Pure forward or backward differences would lead to significantly higher influence of the step size.
As a result of this finding, the step size was chosen for the Monte Carlo derivatives with 0.1 in the following calculations.

Second, the influence of the sample number was investigated as shown in Table~\ref{linear_mcs_results2}.
Similarly as the accuracy of the estimated reliability index, the estimates of the sensitivity indices becomes more accurate with an increasing number of samples. We can observe the same well-known phenomena, that the standard deviation of the estimator scales with the square root of the samples size. However a certain number of samples is necessary, to obtain reliable sensitivity estimates, but the mean of the estimates converges quite well to the reference solution, which clarifies the unbiasedness of the estimation approach.

\begin{table}[th]
\scriptsize
\begin{center}
\begin{tabular}{ccccccc}
\hline
& $\hat \beta$ & $\hat S_{U_1}$& $\hat S_{U_2}$& $\hat S_{U_3}$& $\hat S_{U_4}$& $\hat S_{U_5}$ \rule[0mm]{0pt}{2.5ex}\\[3pt]
   \hline
   Reference	& 3.000 &0.640 &0.250 &0.090 & 0.010 & 0.010  \rule[0mm]{0pt}{2.5ex} \\ 
   \hline
   $N = 100$	& &&&&   \rule[0mm]{0pt}{2.5ex} \\ 
	Mean values	&3.007 &0.626&    0.240&    0.088&    0.023&    0.023\\
	Std deviation	&0.055 &0.055 &   0.049 &   0.040&    0.019 &   0.018\\
   \hline
   $N = 1000$	& &&&&   \rule[0mm]{0pt}{2.5ex} \\ 
	Mean values	&3.000& 0.641&    0.249&    0.089&    0.010&    0.011\\
	Std deviation	&0.018& 0.018&    0.017 &    0.012&    0.008&    0.008\\
   \hline
   $N = 10000$	& &&&&   \rule[0mm]{0pt}{2.5ex} \\ 
	Mean values	&3.000 &0.640&    0.250&    0.089&    0.010&    0.010\\
	Std deviation	&0.005 & 0.005&    0.005&    0.004&    0.003&    0.003\\
   \hline
\end{tabular}
\end{center}
\caption{Linear example: influence of the sample size on the accuracy of the importance sampling sensitivity indices}
\label{linear_is_results}
\scriptsize
\begin{center}
\begin{tabular}{ccccccc}
\hline
& $\hat \beta$ & $\hat S_{U_1}$& $\hat S_{U_2}$& $\hat S_{U_3}$& $\hat S_{U_4}$& $\hat S_{U_5}$ \rule[0mm]{0pt}{2.5ex}\\[3pt]
   \hline
   Reference	&  &0.640 &0.250 &0.090 & 0.010 & 0.010  \rule[0mm]{0pt}{2.5ex} \\ 
   \hline
   $b = 2.0$	& &&&&   \rule[0mm]{0pt}{2.5ex} \\ 
	Mean values	&2.001 &0.633&    0.249&    0.089&    0.016&    0.014\\
	Std deviation	&0.018 & 0.022&    0.019&    0.017&    0.011&    0.010\\
   \hline
   $b = 3.0$	& &&&&   \rule[0mm]{0pt}{2.5ex} \\ 
	Mean values	&3.000& 0.641&    0.249&    0.089&    0.010&    0.011\\
	Std deviation	&0.018& 0.018&    0.017 &    0.012&    0.008&    0.008\\
   \hline
   $b = 4.5$	& &&&&   \rule[0mm]{0pt}{2.5ex} \\ 
	Mean values	&4.502 &0.641&    0.249&    0.091&    0.009&    0.011\\
	Std deviation	&0.016&0.015&    0.013&    0.010&    0.005&    0.005\\
   \hline
   $b = 6.0$	& &&&&   \rule[0mm]{0pt}{2.5ex} \\ 
	Mean values	&6.000 & 0.640&    0.251&    0.088&    0.011&    0.010\\
	Std deviation	&0.014 &0.013&    0.012&    0.009&    0.004&    0.004\\
   \hline
\end{tabular}
\end{center}
\caption{Linear example: influence of reliability index on the accuracy of the importance sampling sensitivity indices using 1000 samples}
\label{linear_is_results2}
\end{table}

Finally, the importance sampling approach was applied.
In table~\ref{linear_is_results} the results are shown depending on the number of samples. Similarly as for the Monte Carlo Simulation, we observed a unbiased convergence of the estimated sensitivity indices with increasing sample size.
Additionally, the influence of the reliability index on the accuracy of the estimated sensitivity measures is analyzed. Table~\ref{linear_is_results2} shows the results,
which clearly indicate, that with decreasing failure probability the proposed approach  
becomes slightly more accurate for the same number of samples.

\subsection{Bearing failure of a shallow foundation}

\begin{figure}[th]
\center
	\includegraphics[width=0.45\textwidth]{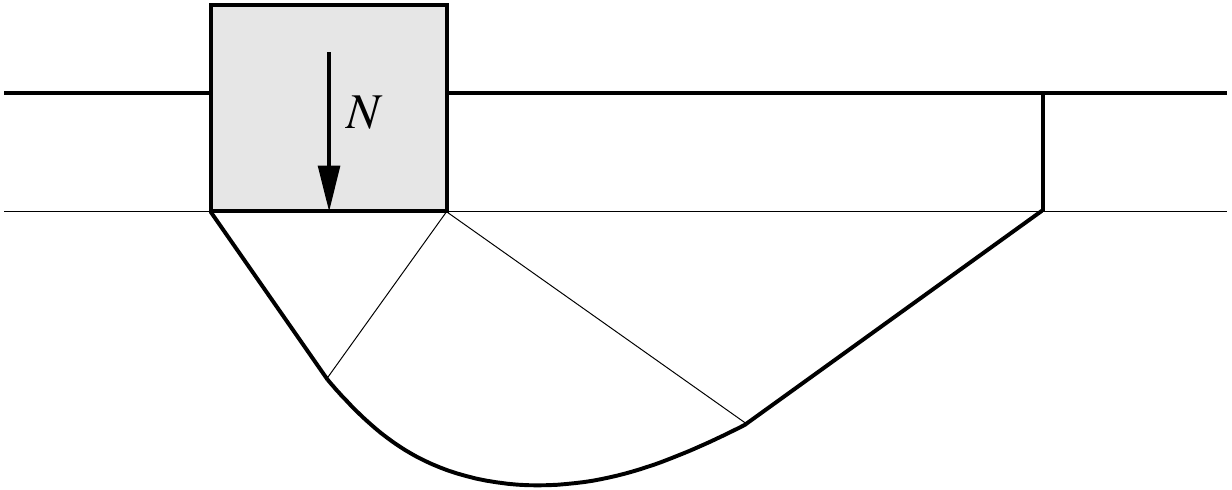}
\scriptsize
\begin{center}
\begin{tabular}{cccc}
\hline
Parameter & Distribution & Mean value & Stand. deviation \rule[0mm]{0pt}{2.5ex}\\[3pt]
   \hline
   Load $N$	& Normal & $200 \,\text{kN}$ & $60 \,\text{kN}$ \rule[0mm]{0pt}{2.5ex} \\ 
	Friction angle $\varphi$ & Log-normal & $20.0^\circ$ &$4.0^\circ$\\
	Cohesion $c$ & Log-normal & $40 \,\text{kN/m}^2$& $12 \,\text{kN/m}^2$\\
	Soil weight $\gamma_s$ & Log-normal & $18 \,\text{kN/m}^3$& $1.8 \,\text{kN/m}^3$\\
 	Width $b$ & Deterministic & $1.5 \,\text{m}$ &-\\
	Depth $d$ & Deterministic & $1.0 \,\text{m}$&-\\
  \hline
\end{tabular}
\end{center}
\caption{Investigated shallow foundation with geometry and soil parameters}
\label{bearing_parameter}
\end{figure}
In the final example, the bearing failure of a shallow foundation is investigated,
in figure~\ref{bearing_parameter} the geometry and the random input parameters are given.

The load bearing capacity can be calculated according to \shortcite{Terzaghi1943} as
\begin{equation}
\begin{aligned}
	R_{sp} &= b\cdot \left(\gamma_s \cdot d \cdot N_{d0} + \gamma_s \cdot b\cdot N_{b0} + c\cdot N_{c0} \right),\\
	N_{d0} &= \tan^2 \left(45^\circ + \frac{\varphi}{2} \right)\cdot e^{\pi \cdot \tan \varphi},\\
	N_{b0} &= \left(N_{d0}- 1\right)\tan \varphi,\\
	N_{c0} &= \frac{N_{d0}-1}{\tan \varphi}.
\end{aligned}
\end{equation}	
Further details about this example can be found in \shortcite{Most_2010_CompGeo}.

Again the First Order Reliability Method, Monte Carlo Simulation and Importance sampling using the FORM design point as
sampling center have been investigated. The reliability sensitivity measures are estimated in the standardized normal space 
similar as the FORM-$\alpha$-factors. In table~\ref{bearing_results} the resulting estimates are given.
Similar as in the previous example, we observed a very good agreement of FORM, MCS and importance sampling estimates.
\begin{table}[th]
\scriptsize
\begin{center}
\begin{tabular}{lccccc}
\hline
& $\hat \beta$ & $\hat S_{U_N}$& $\hat S_{U_\varphi}$& $\hat S_{U_c}$& $\hat S_{U_\gamma}$ \rule[0mm]{0pt}{2.5ex}\\[3pt]
   \hline
   FORM	& 4.31 &0.291 &   0.292  &  0.409 &   0.008 \rule[0mm]{0pt}{2.5ex} \\ 
   \hline
   MCS, $10^7$ samples	& &&&   \rule[0mm]{0pt}{2.5ex} \\ 
	Mean values	  &4.404 & 0.300 &   0.287 &   0.409 &   0.008\\
	Std deviation	&0.025 & 0.027 &   0.023 &   0.027 &   0.009\\
   \hline
   IS, $100$ samples	& &&&   \rule[0mm]{0pt}{2.5ex} \\ 
	Mean values	  &4.406 & 0.293  &  0.287  &  0.405  &  0.015\\
	Std deviation	&0.049 & 0.040  &  0.034  &  0.039  &  0.013\\
   \hline
   IS, $1000$ samples	& &&&   \rule[0mm]{0pt}{2.5ex} \\ 
	Mean values	  &4.401 & 0.295 &   0.291  &  0.407 &   0.008\\
	Std deviation	&0.016 & 0.012 &   0.010  &  0.013 &   0.005\\
   \hline
   IS, $10000$ samples	& &&&   \rule[0mm]{0pt}{2.5ex} \\ 
	Mean values	  &4.401 & 0.294 & 0.290 & 0.410 & 0.006\\
	Std deviation	&0.005 & 0.004 & 0.004 & 0.004 & 0.002\\
   \hline
\end{tabular}
\end{center}
\caption{Bearing failure example: estimated sensitivity indices using FORM, Monte Carlo Simulation (MCS) and importance sampling (IS)}
\label{bearing_results}
\end{table}

\begin{figure}[th]
\center
	\includegraphics[width=0.4\textwidth]{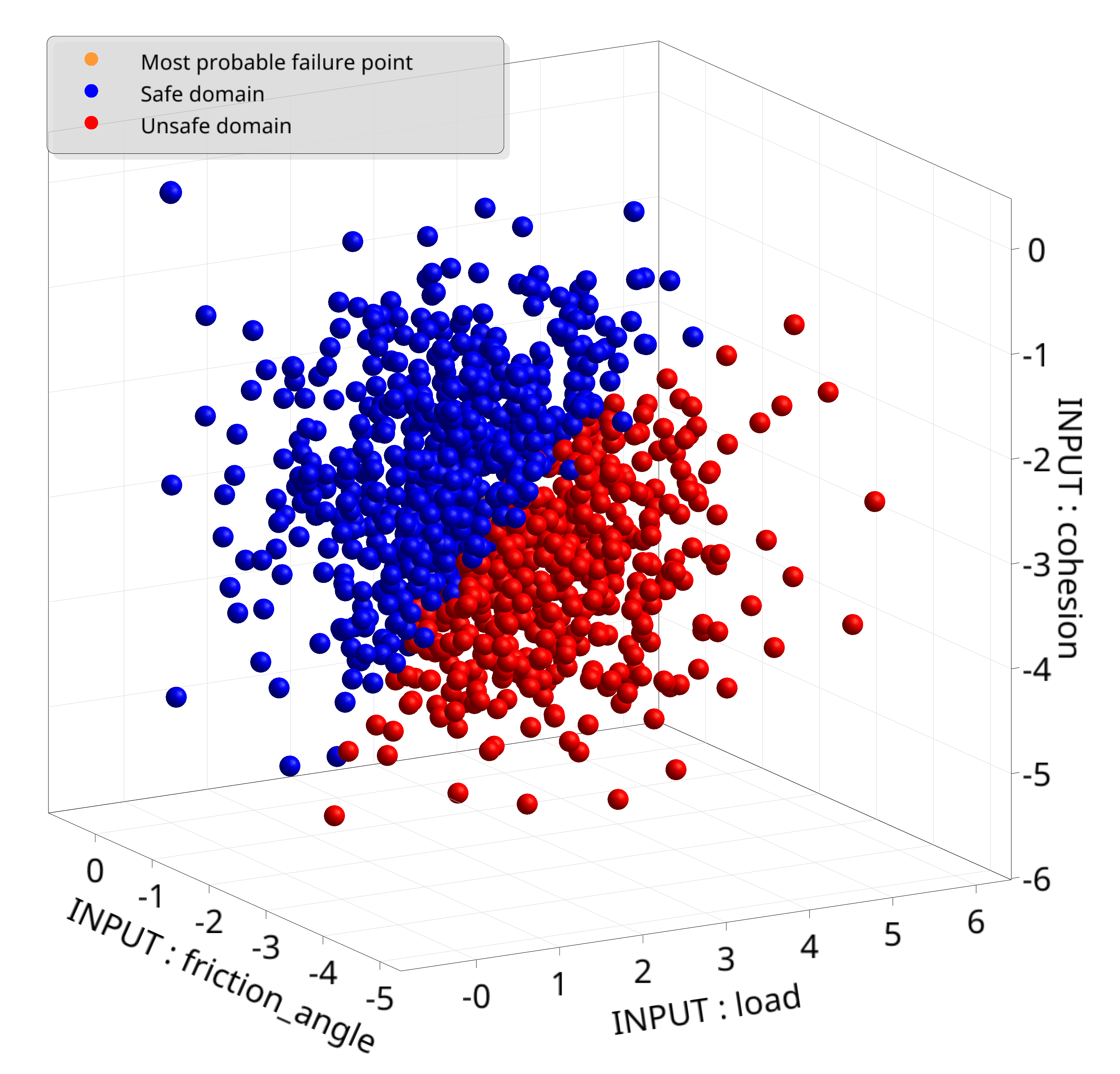}
\caption{Investigated shallow foundation with geometry and soil parameters}
\label{bearing_ispud}
\end{figure}

In figure~\ref{bearing_ispud} the importance sampling samples are shown in the standard normal space for the three most important parameters. The figure indicates, that the limit state function is slightly nonlinear around the most probable failure point,
which explains the observed differences in the  estimates of the reliability index of the FORM and importance sampling approaches.

\section{Conclusions}
In the presented paper a variance-based sensitivity measure was introduced. This measure directly quantifies the influence of the variance of an input parameter with respect to the failure probability. Using an importance sampling weighting for Monte Carlo type methods, these derivatives can be estimated without additional model evaluations just from the existing simulation runs. Thus, these reliability based sensitivity measures can be estimated for every standard or variance-reduced Monte Carlo method within a computational fast evaluation step and can serve the analyst directly an assessment of the input parameter importance.


\end{document}